\documentclass[english]{article}
\usepackage[T1]{fontenc}
\usepackage[latin9]{inputenc}
\usepackage{geometry}
\usepackage{color}
\usepackage{amsbsy}
\usepackage{amstext}
\usepackage[numbers]{natbib}

\makeatletter
\newcommand{\lyxaddress}[1]{
	\par {\raggedright #1
	\vspace{1.4em}
	\noindent\par}
}

\makeatother

\usepackage{babel}
\begin{document}
\title{Waves on shearing current for an arbitrary inertial viewer}
\author{David M. Kouskoulas\thanks{dkouskoulas@mail.tau.ac.il}$\;\;\;$and
Yaron Toledo\thanks{toledo@tau.ac.il}}
\maketitle

\lyxaddress{Marine Engineering and Physics Laboratory, School of Mechanical Engineering,
Tel Aviv University, Tel Aviv 6997801, Israel}
\begin{abstract}
Real world water waves often propagate on current. And, the measurement
of waves and current is an important task for coastal and marine engineers.
Modern marine measurement technologies (i.e. unmanned autonomous vehicles,
drones) often propagate with different velocities relative to the
current and waves. It has been shown that, due to a loss of Galilean
invariance, viewer velocity has a non-trivial effect on the mathematical
form of the wave and uniform current problem. It is demonstrated herein
that similar complexities arise for shearing currents. The work provides
a generalized formulation of the wave and shearing current boundary
value problem for an arbitrary inertial viewer. The moving viewer
dispersion relation for the special case of a constant current shear
is also derived.
\end{abstract}

\section{Introduction}

Real world water waves often propagate on current. The measurement
of waves and current is an important task for coastal and marine engineers.
Such measurements are also relevant to general geophysical research,
e.g. understanding sea surface hydrodynamics and climate. Modern marine
measurement technologies (e.g.. unmanned autonomous vehicles, drones)
often propagate with arbitrary velocities relative to the current
and waves. It has been shown, for waves on uniform current, invariance
between descriptions given by different inertial viewers is not generally
preserved by Galilean transformations \citep{kouskoulas2020linear}.
This is, in part, due to the non-Galilean invariance of Bernoulli's
equation (see also \citep{mungan2011bernoulli}). In real world environments,
it is common to have significant variation of current with depth.
In such cases, uniform current cannot reasonably be assumed.

Water waves have modal structures which are modified by horizontal
and vertical changes in current. Amplitude changes of waves on non-homogeneous
moving media and slowly varying wave guides have been discussed \citep{bretherton1968propagation,bretherton1968wavetrains}.
Effects of horizontal variations on internal waves have been investigated
using the geometric optics approximation \citep{samodurov1974internal,voronovich1976propagation}.
The geometric optics approximation has also been used to derive an
adiabatic invariant for waves on horizontally and vertically changing
flows \citep{voronovich1976close}. The lowest order approximation
yields a governing equation for the vertical structure of the fluid
velocity which is equivalent to Rayleigh's equation or the inviscid
Orr-Sommerfeld equation \citep{skop1987approach,skop1987approximate}.
There is no general solution for the linear theory of waves on arbitrary
shearing current. However, approximate dispersion relations have been
found and demonstrated for certain profiles\textcolor{black}{{} \citep{kirby1989surface,stewart1974hf}.}
This approach was extended to provide an approximated form of the
Wave Action Equation for horizontally inhomogeneous currents on vertical
shear \citep{quinn2017explicit}. Although the effects of vertically
inhomogeneous current on a linear monochromatic wavetrain are reasonably
well understood, there is no discussion regarding how viewer motion
may modify the mathematical form of the problem. This issue has been
overlooked because it is generally assumed that the problem is Galilean
invariant. Since a breaking of Galilean invariance was shown for the
uniform current case \citep{kouskoulas2020linear}, it is of important
to investigate the effects of viewer velocity on the shearing current
problem. This is of particular interest since it is becoming more
common to measure wave-current fields using instrumentation on moving
platforms, both above and below the water surface.

In the following, the surface water wave on shearing current problem
is rewritten for an arbitrary inertial viewer. The boundary value
problem is transformed using Galilean transformations for both coordinates
and derivatives. In a most general case, viewer velocity may be assumed
to have nonzero components in all three spatial dimensions. Equations
of motion governing the vertical wave velocity profile are derived.
With regards to the governing equation, the effects of the viewer
velocity appear in the usual coefficients and as additional derivatives
of the independent variable, increasing the order of the differential
equation. The classical Rayleigh equation is recovered for a fixed
viewer. The kinematic free surface boundary condition is also modified
by the viewer velocity. The effect of horizontal velocity components
appear in the usual coefficients. Similar to the governing equation,
a vertical velocity component increases the order of the differential
equation. There is no general analytical solution for waves on shearing
current. Nonetheless, analytical solutions can be found for certain
special cases. As an example, the case of a constant current shear
is considered. An analytical form of the dispersion relation is derived.
It is shown to reduce to known forms in deep water for uniform and
zero current.

\section{Equations of motion}

Consider incompressible, inviscid and rotational follow and a viewer
denoted by $S$. $\boldsymbol{u}=\left\{ u,v,w\right\} $ is the flow
velocity components, $p$ is the pressure and $\eta$ is the free
surface. Momentum conservation is given by Euler's equations,
\begin{eqnarray}
\frac{\partial\boldsymbol{u}}{\partial t}+\boldsymbol{u}\cdot\nabla\boldsymbol{u}+\frac{1}{\rho}\nabla p & = & 0.\label{eq: Momentum_0}
\end{eqnarray}
Wherein, $\nabla\equiv\left\{ \partial_{x},\partial_{y},\partial_{z}\right\} $.
Equation (\ref{eq: Momentum_0}) follows from the full Navier-Stokes
equations in the absence of viscous effects. Mass conservation is
given by the continuity equation,

\begin{eqnarray}
\nabla\cdot\boldsymbol{u} & = & 0.\label{eq: Continuity_0}
\end{eqnarray}
The free surface boundary conditions are

\begin{eqnarray}
\frac{\partial\mathbf{\eta}}{\partial t}+\boldsymbol{u}\cdot\nabla\eta-w & = & 0,\qquad z=0,\label{eq: KFSBC_0}\\
p-\rho g\eta & = & 0,\qquad z=0,\label{eq: DSBC_0}\\
w & = & 0,\qquad z=-h,\label{eq: BBC_0}\\
w & = & 1,\qquad z=0.\label{eq: FSBC_0}
\end{eqnarray}
Equation (\ref{eq: KFSBC_0}) is the kinematic free surface boundary
condition, (\ref{eq: DSBC_0}) is the dynamic free surface boundary
condition, (\ref{eq: BBC_0}) is the kinematic condition for an impermeable
flat bottom and (\ref{eq: FSBC_0}) normalizes the amplitude of the
harmonic.

Decompose the flow into temporal mean (current) and oscillatory (wave)
components,

\begin{eqnarray}
\boldsymbol{u}\left(\boldsymbol{x},t\right) & = & \boldsymbol{U}\left(z\right)+\epsilon\tilde{\boldsymbol{u}}\left(\boldsymbol{x},t\right),\label{eq: u with perturbation}\\
\boldsymbol{\eta}\left(\boldsymbol{x},t\right) & = & \epsilon\tilde{\eta}\left(x,t\right),\label{eq: eta perturbation}\\
p\left(\boldsymbol{x},t\right) & = & constant+\epsilon\tilde{p}\left(\boldsymbol{x},t\right).\label{eq: p with perturbation}
\end{eqnarray}
Wherein, $\boldsymbol{U}=\left\{ U_{x}\left(z\right),U_{y}\left(z\right),0\right\} $
is the horizontal current. Tildes denote small oscillatory perturbations
corresponding to Fourier components of the wave motion. Substitute
(\ref{eq: u with perturbation})-(\ref{eq: p with perturbation})
into (\ref{eq: Momentum_0})-(\ref{eq: FSBC_0}). The linearized wave
motion is governed by terms $\mathcal{O}\left(\epsilon\right)$,

\begin{eqnarray}
\left(\frac{\partial}{\partial t}+\boldsymbol{U}\cdot\nabla\right)\tilde{\boldsymbol{u}}+w\frac{d\boldsymbol{U}}{dz} & = & -\nabla\tilde{p},\label{eq: Euler_0b}\\
\nabla\cdot\tilde{\boldsymbol{u}} & = & 0.\label{eq: Cont_0b}
\end{eqnarray}
The boundary conditions become

\begin{eqnarray}
\left(\frac{\partial}{\partial t}+\boldsymbol{U}\cdot\nabla\right)\tilde{\eta}-w & = & 0,\qquad z=0,\label{eq: KFSBC_0b}\\
p-\rho g\tilde{\eta} & = & 0,\qquad z=0,\label{eq: DFSBC_0b}\\
w & = & 0,\qquad z=-h,\label{eq: BBC_0b}\\
w & = & 1,\qquad z=0.\label{eq:  FS_0b}
\end{eqnarray}
The mathematical form of equations (\ref{eq: Euler_0b})-(\ref{eq:  FS_0b})
were derived for a fixed viewer who sees the fluid (in the absence
of wave motion) move with the velocity $\boldsymbol{U}$. In a real
world context, this viewer would correspond to a measurement instrumentation
attached to a fixed platform relative to the earth (e.g. wave gauge
attached to sea floor in shallow water). However, modern measurement
technologies may also move with arbitrary velocities and directions
relative to the earth (and waves and current). For example, they may
be ship-towed, attached to unmanned autonomous vehicles or deployed
on a moving drone. It is a common assumption that (\ref{eq: Euler_0b})-(\ref{eq:  FS_0b})
will have the same form for all inertial viewers. However, it will
be shown that this is not generally the case within a Galilean framework
since the system exhibits broken Galilean boost invariance.

\section{System in an inertial reference frame}

Assume an inertial reference frame $S'$ moving with a constant velocity
$\boldsymbol{V}=\left\{ V_{x},V_{y},V_{z}\right\} $ relative to $S$.
In the Galilean framework, invariance between inertial reference frames
is given by the following transformations,

\begin{equation}
\begin{array}{c}
\begin{array}{c}
\boldsymbol{x}=\boldsymbol{x}'+\boldsymbol{V}t',\hphantom{her}t=t',\hphantom{hereisa}\end{array}\\
\nabla=\nabla',\hphantom{her}\partial_{t}=\partial{}_{t'}-\boldsymbol{V}\cdot\nabla'.
\end{array}\label{eq: coord. transformations}
\end{equation}
These are the Galilean transformations. In $S'$, Euler's equation
and the continuity equation become

\begin{eqnarray}
\left(\frac{\partial}{\partial t'}+\left(\boldsymbol{U}-\boldsymbol{V}\right)\cdot\nabla'\right)\tilde{\boldsymbol{u}}'+\tilde{w}'\frac{d\boldsymbol{U}}{dz'} & = & -\nabla'\tilde{p}',\label{eq: Euler_1}\\
\nabla'\cdot\tilde{\boldsymbol{u}}' & = & 0.\label{eq: Cont_1}
\end{eqnarray}
The boundary conditions take the form,

\begin{eqnarray}
\left(\frac{\partial}{\partial t'}+\left(\boldsymbol{U}-\boldsymbol{V}\right)\cdot\nabla'\right)\tilde{\eta}'-\tilde{w}' & = & 0,\qquad z'=0,\label{eq: KFSBC_1}\\
\tilde{p}'-\rho g\tilde{\eta}' & = & 0,\qquad z'=0,\label{eq: DFSBC_1}\\
\tilde{w}' & = & 0,\qquad z'=-h,\label{eq: BBC_1}\\
\tilde{w}' & = & 1,\qquad z'=0.\label{eq:  FS_1}
\end{eqnarray}
Primes denote quantities in $S'$. For clarity in writing, the primes
will now be dropped. Introducing temporal and spatial periodicity
into the oscillatory terms yields,

\begin{eqnarray}
\tilde{\boldsymbol{u}}\left(\boldsymbol{x},t\right) & = & \hat{\boldsymbol{u}}\left(z\right)\exp\left[i\left(k_{x}x+k_{y}y-\left(-1\right)^{n}\omega_{d}t\right)\right],\label{eq: u oscillatory}\\
\tilde{\boldsymbol{\eta}}\left(\boldsymbol{x},t\right) & = & \hat{\eta}\exp\left[i\left(k_{x}x+k_{y}y-\left(-1\right)^{n}\omega_{d}t\right)\right],\label{eq: eta oscillaotry}\\
\tilde{\boldsymbol{p}}\left(\boldsymbol{x},t\right) & = & \hat{p}\left(z\right)\exp\left[i\left(k_{x}x+k_{y}y-\left(-1\right)^{n}\omega_{d}t\right)\right].\label{eq: p oscillatory}
\end{eqnarray}
Wherein, $n=1,2$ differentiate between solution branches (see appendix
C, \citep{kouskoulas2020linear}), $\hat{\boldsymbol{u}}=\left\{ \hat{u},\hat{v},\hat{w}\right\} $,
and $\omega_{d}=\omega-\boldsymbol{k}\cdot\boldsymbol{V}$ is the
Doppler shifted frequency ($\omega$ is the frequency measured in
$S$). 

Combining (\ref{eq: Euler_1}) and (\ref{eq: Cont_1}), using the
standard derivation procedure for a fixed viewer (see, for example,
\citep{voronovich1976close}), yields a governing equation in terms
of $\hat{w}$ in $S'$,

\begin{equation}
\frac{\partial^{2}\hat{w}}{\mathcal{\partial}z^{2}}-\left(\left|\boldsymbol{k}\right|^{2}+\frac{1}{(-1)^{n+1}\mathcal{C}-\mathcal{U}+\mathcal{V}}\frac{\partial^{2}\mathcal{U}}{\mathcal{\partial}z^{2}}\right)\hat{w}+\frac{iV_{z}}{\left|\boldsymbol{k}\right|\left((-1)^{n+1}\mathcal{C}-\mathcal{U}+\mathcal{V}\right)}\left(\left|\boldsymbol{k}\right|^{2}\frac{\partial\hat{w}}{\partial z}-\frac{\partial^{3}\hat{w}}{\partial z^{3}}\right)=0.\label{eq: gov combined}
\end{equation}
Wherein, the following quantities have been introduced,

\begin{equation}
\mathcal{U}=\frac{\boldsymbol{k}\cdot\boldsymbol{U}}{\left|\boldsymbol{k}\right|},\qquad\mathcal{V}=\frac{\boldsymbol{k}\cdot\left\{ V_{x},V_{y}\right\} }{\left|\boldsymbol{k}\right|},\qquad\mathcal{C}=\frac{\omega_{d}}{\left|\boldsymbol{k}\right|}.\label{eq:}
\end{equation}
$\mathcal{U}$ and $\mathcal{V}$ are the projections of the current
and viewer velocities on the wave. $\mathcal{C}$ is the wave celerity.
As expected, for $\left\{ V_{x},V_{y}\right\} =V_{z}=0$ (a fixed
viewer), Rayleigh's equation is recovered. It is of interest to inspect
the influence of different viewer velocity components on (\ref{eq: gov combined}).
When $V_{x}\neq0$ and $V_{y}\neq0$ and $V_{z}=0$, the general functional
form of Rayleigh's equation is preserved; however, additional secularities
appear in the coefficients. Perhaps more problematic is the case of
a viewer velocity with a vertical component, i.e. $V_{z}\neq0$. This
introduces an imaginary term in (\ref{eq: gov combined}) which has
no analogy in $S$. Moreover, it introduces a third order derivative,
increasing the order of the differential equation governing the independent
variable $w$. This renders the boundary value problem under-determined.
An additional (possibly symmetry) condition will be required in order
to uniquely determine the solution.

Combining (\ref{eq: DFSBC_1}) and (\ref{eq: Euler_1}) and eliminating
$\hat{\boldsymbol{u}}$ $vis$ $a$ $vis$ (\ref{eq: Cont_1}) yields
a combined free surface boundary condition in $S'$,

\begin{equation}
\frac{\partial\hat{w}}{\partial z}+\left(\frac{1}{\left((-1)^{n+1}\mathcal{C}-\mathcal{U}+\mathcal{V}\right)}\frac{\partial\mathcal{U}}{\mathcal{\partial}z}-\frac{g}{\left((-1)^{n+1}\mathcal{C}-\mathcal{U}+\mathcal{V}\right)^{2}}\right)\hat{w}-\frac{iV_{z}}{\boldsymbol{k}\left((-1)^{n+1}\mathcal{C}-\mathcal{U}+\mathcal{V}\right)}\frac{\partial^{2}\hat{w}}{\partial z^{2}}=0.\label{eq: CFSBC S'}
\end{equation}
As expected, setting $\left\{ V_{x},V_{y}\right\} =V_{z}=0$ recovers
a known form of the boundary condition for a fixed viewer. Similar
to (\ref{eq: gov combined}), when $V_{z}\neq0$ an imaginary term
without analogy in $S$ appears. Therefore, the system exhibits a
loss of Galilean boost invariance for a viewer moving with a vertical
velocity component (e.g. an underwater autonomous vehicle changing
depth).

\section{Approximate dispersion relation: constant shear}

There is no general analytical solution to (\ref{eq: gov combined})-(\ref{eq: CFSBC S'}).
However, for certain current profiles and viewers, one may find analytical
forms of dispersion. Consider here the simple case of a constant shear
and viewer moving in the horizontal plane ($V_{x}\neq0$, $V_{y}\neq0$,
and $V_{z}=0$). The current profile will have the form,

\begin{equation}
\mathcal{U}\left(z\right)=\text{\ensuremath{\mathcal{A}+\mathcal{B}z}}.
\end{equation}
The dispersion relation becomes,

\begin{eqnarray}
\mathcal{C} & = & \frac{(-1)^{-3n}\left(-2\boldsymbol{k}\left(\mathcal{A}-\mathcal{V}\right)+\left(\mathcal{B}+\left(-1\right)^{m}\sqrt{\mathcal{B}^{2}+4g\boldsymbol{k}\coth(\boldsymbol{k}h)}\right)\tanh(\boldsymbol{k}h)\right)}{2\boldsymbol{k}}.\label{eq: C gen}
\end{eqnarray}
Wherein, $m=1,2$ and $n=1,2$. In deep water, $\boldsymbol{k}h\rightarrow\infty$,
(\ref{eq: C gen}) becomes

\begin{eqnarray}
\mathcal{C} & = & \frac{(-1)^{-3n}\left(-2\boldsymbol{k}\left(\mathcal{A}-\mathcal{V}\right)+\left(\mathcal{B}+\left(-1\right)^{m}\sqrt{\mathcal{B}^{2}+4g\boldsymbol{k}}\right)\right)}{2\boldsymbol{k}}.\label{eq: C_deep}
\end{eqnarray}
Known forms can be recovered for a fixed viewer. Consider two special
cases: uniform and zero current in deep water.

\noindent $\mathbf{Case\;1:}$ Fixed viewer of wave propagation on
uniform current in deep water $\left(\mathcal{\mathcal{B}}\rightarrow0,\boldsymbol{k}h\rightarrow\infty\right)$,

\begin{eqnarray}
\mathcal{C} & = & \frac{(-1)^{-3n}\left(-\boldsymbol{k}\mathcal{A}+\left(-1\right)^{m}\sqrt{g\boldsymbol{k}}\right)}{\boldsymbol{k}}.\label{eq: fixed_uniform}
\end{eqnarray}

\noindent $\mathbf{Case\;2:}$ Fixed viewer of wave propagation on
still and deep water $\left(\mathcal{\mathcal{A}}\rightarrow0,\mathcal{\mathcal{B}}\rightarrow0,\boldsymbol{k}h\rightarrow\infty\right)$,

\begin{equation}
\mathcal{C}=\frac{(-1)^{-3n+m}\sqrt{g\boldsymbol{k}}}{\boldsymbol{k}}.\label{eq: disp still and deep}
\end{equation}
Thus, the dispersion relation, (\ref{eq: C gen}), is consistent with
known approximations in the limits of deep water with uniform and
zero current.

\section{Discussion}

The linear gravity wave on shearing current problem was generalized
for an arbitrary inertial viewer using Galilean transformations. Results
demonstrate that the functional form of the boundary value problem
is not the same for all inertial viewers. Thus, Galilean invariance
applies only in a restricted sense. The effect of a vertical viewer
velocity component is of particular interest. In this case, imaginary
terms appear in the governing equation and free surface condition
which have no analogy for a fixed viewer. It should be noted that
the effect remains in the limit of zero current. If the viewer velocity
is strictly in the horizontal plane, there is no imaginary term; however,
new secularities appear in the usual coefficients.

There is no general analytical solution for wave propagation on an
arbitrary shearing current profile. Nonetheless, for some simple cases
analytical solutions may be found. As an example, the dispersion relation
for a moving viewer of waves on current with constant shear was derived.
For simplicity, it was assumed the viewer moves only along a horizontal
plane. This avoids the imaginary terms which are generated by a vertical
component in the viewer velocity. The new generalized dispersion relation
was shown to recover known forms in the limits of uniform and zero
current in deep water.

Wave measurement instrumentations often propagate with different directions
and velocities relative to the waves and current. And, in most real
world environments, there will be vertical variation in a current
profile. Using a Galilean framework, the preceding results demonstrate
the combined effect of a moving viewer and current shear on the functional
form of dispersion. The results are of high importance to the interpretation
of wave and current measurements by moving platforms.

\bibliographystyle{plainnm}
\bibliography{ShearMoving}

\end{document}